\documentclass[12pt,preprint]{aastex}

\shorttitle{Black Hole Formation in Clumpy Disks}

\shortauthors{Elmegreen, Bournaud, Elmegreen}

\begin{document}

%\title{Formation of Nuclear Black Holes from the Coalescence of Giant Star
%Forming Clumps in Primordial Disk Galaxies }

\title{Nuclear Black Hole Formation in Clumpy Galaxies at High Redshift}

\author{Bruce G. Elmegreen}
\affil{IBM Research Division, T.J. Watson Research Center, P.O. Box
218, Yorktown Heights, NY 10598, USA} \email{bge@watson.ibm.com}
\author{Fr\'ed\'eric Bournaud}
\affil{Laboratoire AIM, CEA-Saclay DSM/IRFU/SAP - CNRS - Universit\'e
Paris Diderot, F-91191~Gif-sur-Yvette Cedex, France}
\email{frederic.bournaud@cea.fr}
\author{Debra Meloy Elmegreen}
\affil{Vassar College, Dept. of Physics \& Astronomy, Box 745,
Poughkeepsie, NY 12604} \email{elmegreen@vassar.edu}

\begin{abstract}
Massive stellar clumps in high redshift galaxies interact and migrate
to the center to form a bulge and exponential disk in $\lesssim1$ Gyr.
Here we consider the fate of intermediate mass black holes (BHs) that
might form by massive-star coalescence in the dense young clusters of
these disk clumps. We find that the BHs move inward with the clumps and
reach the inner few hundred parsecs in only a few orbit times. There
they could merge into a supermassive BH by dynamical friction. The
ratio of BH mass to stellar mass in the disk clumps is approximately
preserved in the final ratio of BH to bulge mass. Because this ratio
for individual clusters has been estimated to be $\sim10^{-3}$, the
observed BH-to-bulge mass ratio results. We also obtain a relation
between BH mass and bulge velocity dispersion that is compatible with
observations of present-day galaxies.

\end{abstract}


\section{Introduction}
\label{sect:intro}

Numerical simulations have reproduced the massive clumpy structures of
star formation in high-redshift galaxies and followed the migration of
these clumps into the galaxy centers where they merge to form bulges
(Noguchi 1999; Immeli 2004ab; for a review of clumpy structures, see
Elmegreen 2007). The clumps result from gravitational instabilities in
a gas-rich, highly-turbulent disk, and the central migration results
from clump interactions and angular momentum losses to the disk, halo,
and clump debris.  In a series of papers, we have shown that the
resulting disk has the characteristic double-exponential profile of
modern spiral galaxies (Bournaud, Elmegreen \& Elmegreen 2007,
hereafter BEE07), and the bulge has a classical form, with high Sersic
index, three-dimensional random motions,  little rotation, and a rapid
formation (Elmegreen, Bournaud, \& Elmegreen 2008; hereafter EBE08). We
have also reproduced in detail the peculiar morphology and kinematics
of a galaxy in the Hubble Space Telescope Ultra Deep Field, UDF 6462,
with this model (Bournaud et al. 2008). Other spectroscopic
observations also indicate, less directly but over a larger sample,
that high-redshift disk and bulge evolution is characterized by giant
clump interactions and high turbulence (e.g., F\"orster Schreiber
2006), which is consistent with our models.

This paper considers another aspect of clumpy disk evolution, the
formation of nuclear black holes (BHs). Models of bulge formation
should be able to explain how BHs form at the same time, why the
BH-to-bulge mass ratio is $\simeq 0.002$ (McLure \& Dunlop 2002;
Marconi \& Hunt 2003), and why the BH mass and bulge central velocity
dispersion are related by $\log(M_{BH}/M_\odot)=
8.13+4.02\log(\sigma_{bulge}/200\;{\rm km\;s}^{-1})$ (Ferrarese \&
Merritt 2000; Tremaine et al. 2002). Here we model all of these
observations by considering that each clump forms an intermediate mass
black hole (IMBH) by stellar coalescence. We follow the evolution of
these IMBHs as their clumps move in the disk. We find that the IMBHs
migrate inward along with the clumps and that the final central
BH-to-bulge mass ratio is approximately the same as the initial
BH-to-clump ratio. This is about the observed value for bulges. The
velocity dispersion relation for BHs in bulges also results to a
reasonable approximation.

In the following, Section \ref{sect:model} outlines our model for
nuclear BH formation, section \ref{sect:num} describes the numerical
simulations, section \ref{sect:results} gives the results, and section
\ref{sect:discu} contains a brief discussion.

\section{Black Hole Formation Model}\label{sect:model}

Nuclear BHs are an important aspect of galaxy and bulge formation.
Malbon et al. (2007) summarized BH models by suggesting that gas
accretion during starbursts forms relatively low mass BHs at high
redshift, while BH coalescence during galaxy mergers forms supermassive
BHs at low redshift (see also works by Di~Matteo et al. 2005, 2007;
Johansson, Naab, \& Burkert 2008). Disk density waves are less
efficient in fueling nuclear black holes (Younger et al. 2008). In a
very different type of model, Ebisuzaki et al. (2001) suggested that
IMBHs grow by stellar coalescence in dense young clusters that form in
the central regions of galaxies. Dynamical friction then forces these
IMBHs to the center where they merge into a nuclear BH.  Here we
determine whether a model like this can also apply to IMBHs that form
in dense disk clusters, far from the nucleus. We know from clumpy disk
models that the disk clusters migrate to the center to form a bulge, so
the primary question here is whether IMBHs that form in these clusters
follow them inward to the nucleus. The BH-to-cluster mass ratio was
found in the simulations by Ebisuzaki et al. (2001) to be
$\sim10^{-3}$, which is the same as the BH-to-bulge mass ratio.  Thus,
what we primarily need to determine is whether this mass fraction is
preserved during the clump/BH migration. There are two important
differences from the Ebisuzaki et al. model: (1) the clusters here are
much more massive than they considered, so the IMBHs are more massive
also ($\sim10^5\;M_\odot$ each in our models), and (2) our clusters
start in the main galaxy disk rather than the central regions, so they
come in as a result of gravitational interactions between clumps, not
dynamical friction.  We also assume that the giant clumps observed in
high redshift disks are composed of denser, unresolved clusters, where
the stars actually form, just as star complexes in local galaxies form
most of their stars in dense clusters.  The IMBHs presumably form
inside these dense clusters.

Cluster simulations have generally supported this model for IMBH
formation. Portegies-Zwart \& McMillan (2002) confirmed that runaway
stellar collisions can make cluster BHs with $10^{-3}$ of the cluster
mass during core collapse if the relaxation time is less than 25 Myr.
G\"urkan, Freitag \& Rasio (2004) did a cluster simulation with $10^7$
stars and found mass segregation and core collapse in less than an
O-star lifetime, at which point an IMBH with $10^{-3}$ of the cluster
mass formed. Portegies-Zwart et al. (2004) applied the model to the
suspected IMBH in M82 (Matsumoto et al. 2001). G\"urkan, Fregeau, \&
Rasio (2006) showed that stellar collisions in clusters with binaries
could make two IMBHs which form their own binary. Freitag, G\"urkan \&
Rasio (2006) included stellar collisions in a cluster simulation and
found that an IMBH forms when core collapse happens faster than the
main sequence lifetime of a massive star; this requires densities of
$10^6-10^7$ pc$^{-3}$ (Freitag 2007). For the Milky Way nucleus,
Portegies-Zwart et al. (2006) showed that 10\% of dense clusters form
IMBHs during their inward migration and they coalesce fast enough to
explain the massive BH now. Matsubayashi, Makino, \& Ebisuzaki (2007)
found that dynamical friction on IMBHs becomes ineffective close to the
central BH, but the IMBH spirals in anyway because of gravitational
radiation. These theoretical studies suggest that at least some nuclear
BHs could have formed by the merger of IMBHs from dense massive
clusters. Subsequent gas accretion would presumably have caused the BHs
to grow to their present masses. Feedback effects (e.g., McLaughlin et
al. 2006) might be important at this later gas accretion stage.

\section{Numerical Simulations} \label{sect:num}

The evolution of gas-rich galaxy disks is modeled with a particle-mesh
sticky-particle code (Bournaud \& Combes 2002, 2003) with a grid
resolution and gravitational softening length of 110~pc. Stars, gas,
and dark matter halo are each modeled with one million particles. The
sticky-particle parameters are $\beta_r=\beta_t=0.7$ for all but run 4,
which has $\beta_r=\beta_t=0.8$. We assume a local Schmidt law for star
formation in which the probability per timestep that each gas particle
is transformed into a stellar particle is proportional to the 1.4 power
of the local gas density (Kennicutt 1998). The proportionality factor
gives a star formation rate of $3.5$~M$_{\sun}$~yr$^{-1}$ in the
initial disk. Star formation feedback is not expected to be important
for the formation and lifetime of the clumps because of their enormous
masses and deep local potential wells (BEE07).

The initial model parameters were summarized in Table 1 of BEE07; runs
1-7 are the same here except for the addition of BHs. We review the
assumptions briefly here. The initial disk is composed of gas and stars
with a uniform surface density. This follows from the observation that
extremely clumpy disks at $z\ge1$ do not have exponential profiles
(Elmegreen et al. 2005). The initial disk radius is 6~kpc (typical for
clump cluster, chain, and spiral galaxies at $z\ge1$ -- Elmegreen et
al. 2007a) and the initial thickness is $h=700$~pc with a
$\mathrm{sech}^2\left( z/h \right)$ vertical distribution (typical for
edge-on spiral and chain galaxies at $z\ge1$ -- Elmegreen \& Elmegreen
2006). The disk mass is $7 \times 10^{10}$~M$_{\sun}$. Stars have a
Toomre parameter in the stable regime, $Q_{\mathrm{s}}=1.5$. The
initial velocity dispersion of the gas, $\sigma_{gas}$, is 9 km
s$^{-1}$ for all runs but 1, 2, and 3, where it is 5, 15, and 20 km
s$^{-1}$, respectively. The gas mass fraction in the disk,
$f_{\mathrm{G}}$, is 0.5 except for runs 4 and 5, where it is 0.25 and
0.75. The halo-to-disk mass ratio, $H/D$, inside the initial disk
radius, is 0.5 except for runs 6 and 7, where it is 0.25 and 0.80. The
small number of massive clumps observed in high-redshift disks implies
a relatively high ratio of turbulent speed to rotation speed,
$\sim10$\% or more. To get such a turbulent disk gravitationally
unstable so that it forms clumps, we need a fairly high gas column
density, which means a high gas-to-star ratio in the disk at that time
(see BEE07, EBE08).

We have performed simulations for our series of studies that have three
types of initial halo and bulge properties: runs~0 to 7 have no initial
bulge and a dark halo that is a Plummer sphere with a scale-length of
15~kpc. Runs 0N, 1N, and 2N have a $\Lambda$CDM cuspy halo (Navarro,
Frenk \& White 1997) with a cusp scale-length $r_S = 6$~kpc
(concentration parameter 16.7 for a virial radius of 100~kpc). Runs 0B,
1B, and 2B have a small initial bulge that is a Plummer sphere with
10\% of the disk mass and a radial scale-length of 600~pc. The other
parameters for these six runs, including the gas fractions and
halo-to-disk mass ratios, are unchanged from runs~0, 1, and 2,
respectively.

Massive clumps form quickly in all simulations, in about the local
dynamical time at the midplane gas density. Their masses and sizes are
comparable to the local Jeans mass and size. The clumps were identified
objectively every 25~Myr as regions where the surface density is
locally larger than the radial average by a factor of 3. Only clumps
with masses larger than $2\times10^8$ $M_\odot$ and sizes smaller than
3~kpc were considered to be clumps (this avoids misidentifying spiral
arms as clumps). When the mass fraction in the clumps reached its
maximum value (usually at a time of about 200~Myr in the simulation),
single particles representing BHs were positioned in the centers of all
the identified clumps, one for each clump. The particle masses were
equal to $10^{-3}$ times the clump masses at that time. They were
initially placed at the positions of peak density. To ensure mass
conservation, we removed the corresponding number of gas and star
particles, randomly chosen within the gravitational softening length.
Thereafter, the BH particles were treated like massive star particles
that could be moved only by gravity. The $10^{-3}$ BH fraction in the
clumps was fixed for the reasons detailed above. We do not aim at
resolving the processes driving it, like the mass accretion onto the
BHs and the potentially associated AGN feedback.

The BH particles have such low masses that they do not affect the clump
dynamics, as confirmed by detailed comparisons with runs having no BH
particles. Neither do the BH particles interact with each other much in
the disk environment. If we consider an interaction to be an approach
within 200 pc, then the number of such interactions at radii greater
than 2 kpc in the disk is a total of 6 for all 8 runs and 58 BHs
produced in these runs. That is a 10\% effect. An additional 4 total
BHs interacted between 1 and 2 kpc radius. By far most of the BHs
interact with each other only when they get to the nucleus.

\section{Results} \label{sect:results}

The time evolution of the gas+star mass column density is shown for
run~0 in Figure 1. White dots represent the initial IMBH particles.
They form inside the clumps at moderate to large radii in the
primordial disk, and then migrate to the center with the clump cores.
The entire process takes about 1 Gyr. Run~0 models without BHs were
shown in BEE07 and run~0 models with cuspy dark matter halos (0N, 1N,
2N) and small initial bulges (0B, 1B, 2B) were shown without BHs in
EBE08. The clump evolution is indistinguishable when BHs are included,
so we do not repeat the 0N and 0B figures here in the BH cases.

Table 1 lists the numbers of BHs that formed and the numbers that get
within the central 250, 500, and 1000~pc by the end of each run. It
also gives the ratios of the BH masses that reach the inner 500 pc or
1000 pc to the bulge masses. Some low-mass BHs from low-mass clumps do
not reach the central regions, but most BHs reach the central 500 pc
and even the central 250 pc.  The final bulge and BH masses are also
tabulated, as are the bulge-to-total mass ratios, the BH-to-bulge mass
ratios, and the bulge velocity dispersions. These dispersions,
$\sigma_{bulge}$, are the average line-of-sight values for 50 2-D
projections of the stars, uniformly distributed over the sine of the
inclination angle. Each dispersion comes from all the stars inside a
projected aperture diameter of 110 pc (the grid resolution). We do not
subtract the disk component from edge-on projections, nor do we make a
direct 3-D measurement, because these would not be done in
observations.

Figure \ref{fig:BHrad} plots the final versus the initial BH
galactocentric radii to show how common it is for the clump BHs to
reach the galaxy centers. Black holes that start in clumps that are far
out in the disk do not typically reach the center; they move inward
only about a factor of 2 in radius and then the clump disperses without
adding to the bulge. The figure also includes simulations that have
cuspy halos (circles) and those that have a small initial bulge. In all
cases, clumps form in the disk by gravitational instabilities, and BHs
that are placed in these clumps migrate inward along with the clumps.
Those that start within $\sim4$ kpc get all the way to the center in 1
Gyr or less. The BHs that get in the furthest are the most massive ones
that formed in the most massive clumps (Table 1).

Bulges and centralized BH clusters always form together in our models.
The total mass of the centralized BHs (those reaching the central 500
pc) is usually within a factor of 2 of the initial BH mass formed in
the clumps. The IMBHs that do not reach the central regions (those from
the smallest clumps) represent a small fraction of the total BH mass.
The BH/bulge mass fraction can therefore be written,
\begin{equation}
{{M_{BH}}\over{M_{bulge}}}\sim\left({{M_{BH}}\over{M_{clump}}}\right)
\left({{M_{clump}}\over{M_{bulge,clumps}}}\right)
\left({{M_{bulge,clumps}}\over{M_{bulge}}}\right)
\end{equation}
where $M_{BH}/M_{clump}=10^{-3}$ according to cluster simulations. In
this equation, $M_{clump}$ is the total clump mass, $M_{bulge,clump}$
is the mass in the bulge that comes from the clumps, and $M_{bulge}$ is
the total bulge mass. Typically $M_{clump}/M_{bulge,clumps}\sim2$ and
$M_{bulge,clumps}/M_{bulge}\sim0.5$ (see BEE07). Note that half of the
bulge stars come from intense star formation in the bulge region during
the clump merging process. This makes up for the half of the clump
stars that are left behind in the disk. Gas accretion onto the BHs
should increase their mass after they get to the nucleus, and other
processes could add to the bulge and disk later too.

Figure \ref{fig:disp} shows the final nuclear BH mass, taken from the
mass of IMBHs that migrated within radii $<500$ pc, versus the final
bulge line-of-sight velocity dispersion. There is a clear correlation.
The dashed line has a slope of 4, which is the observed value, and the
solid line has a slope of 2, which is a reasonable approximation to the
lower envelope. The absolute scale depends on uncertain assumptions
about the initial value of $M_{BH}/M_{clump}$, and on whether we assume
nuclear BHs come from IMBHs within 250~pc or 1000~kpc (see Table 1).
For a dispersion of 130 km s$^{-1}$, which is in the middle of the
plot, the Tremaine et al. (2002) BH mass would be
$2.4\times10^7\;M_\odot$, whereas we get half this value,
$1.3\times10^7\;M_\odot$. We consider the existence of a correlation to
suggest that the proposed model is plausible. The model details are too
crude and the bulge mass range considered is too small to be more
conclusive at this time. For example, the BH/clump mass fraction,
assumed to be a constant $10^{-3}$, could be higher in higher mass
clumps or in more massive galaxies; this could change the correlation
slightly. Other processes are likely to contribute to the bulge
dispersions and the BH masses too, and they could alter the slope. For
example, AGN feedback (neglected in our simulations) could bring the
slope of the BH mass - velocity dispersion relation closer to the
presently observed value (McLaughlin et al. 2006) if BH formation
occurs quickly, as in our present model. Observations at high redshift
suggest that the correlation evolves (Peng et al. 2006ab; Treu et al.
2007; Woo et al. 2008).

\section{Discussion}\label{sect:discu}

Numerical simulations suggest that primitive gas-rich disks should
fragment into clumpy star-forming complexes with masses of
$\sim10^8\;M_\odot$ or more, and that these complexes should interact
gravitationally and move to the galaxy center where they combine to
form a bulge. The surrounding disk becomes exponential in the process.
If the complexes contain massive dense clusters, and if these clusters
form IMBHs, then the IMBHs will migrate to the center too. Our model
assumes that the centralized BHs eventually merge into a single,
nuclear BH. This assumption makes our model significantly different
than BH formation models that rely on massive gas accretion to feed a
seed nuclear BH.

Relativistic effects during BH mergers cannot be treated by our model.
Even their capture into BH binaries as a first step toward merging
cannot be simulated, because the gravity softening length is comparable
to their final separation. Still, BH binaries should form because of
strong dynamical friction in a gas-rich environment (Escala et al.
2005). This process has been resolved in models by Mayer et al. (2007).
Gravitational radiation should then drive their eventual merger.
However, coalescing BHs can experience velocity kicks from the
anisotropy of gravity waves, and these velocity kicks can be as large
as hundreds or even thousands of km s$^{-1}$ (e.g., Baker et al. 2007).
Such fast-moving BHs could be ejected from the galaxy (Merritt et al.
2004). This would challenge many models of nuclear BH formation,
including the common idea that they grow from mergers during the
hierarchical galaxy build-up (e.g., Di~Matteo et al. 2007).

A solution to this problem was proposed by Bogdanovic et al. (2007) in
the hierarchical merging context. According to Baker et al. (2007), the
kick is greatly reduced if the merging BHs have spins aligned with each
other. Bogdanovic et al. (2007) showed that such alignments can result
from torques in gas-rich merging galaxies. In the present scenario
where disk clumps merge together, gas torques may contribute in the
same way, but it is also likely that the BHs will have their spins
already aligned with their orbit angular momenta. This is because the
clumps in which they form are all fragments of the same initial galaxy
disk; they have the same spin orientation and the same orbital angular
momentum when they coalesce (BEE07; Bournaud et al. 2008). The BHs
should preserve these alignments and coalesce without experiencing
major kicks. Even clumps that merge outside the central kpc should have
their spins aligned, again reducing the chance for velocity kicks in
BHs that merge. We also note that recoiling BHs would settle back
rapidly to the galaxy center, as long as the velocity kick does not
exceed the escape velocity (Blecha \& Loeb 2008).

Both classical and pseudo-bulges have nuclear BHs with masses
proportional to the bulge mass (Kormendy \& Richstone 1995; Magorrian
et al. 1998) and a power of the velocity dispersion (Ferrarese \&
Merritt 2000; Gebhardt et al. 2000; Novak et al. 2006). Active galaxies
follow the same relation as non-active galaxies (Nelson 2000).  These
correlations may change slightly over time. At intermediate redshifts,
$z\sim 0.36-1$, the bulge mass may be $\sim2\times$ smaller than the
modern bulge mass compared to the BH mass (Treu et al. 2007), and at
higher redshifts, $z>1.7$, the bulge may be $\sim4\times$ smaller (Peng
et al. 2006ab). Thus bulges seem to grow for a slightly longer time
than BHs. Salviander et al. (2007) and Lauer et al. (2007) caution that
some of this appearance of late bulge growth may result from
observational bias in favor of more active nuclei. In our model the
time evolution may be explained by continued bulge growth during minor
mergers and secular evolution, after the first bulge and its BH formed
via clump/IMBH coalescence. If it takes a highly gas-rich disk to make
clusters dense enough to form IMBHs, then only the first generation of
clumps will add to the nuclear BH and the rest may add only to the
bulge.

Small-bulge galaxies ($M<10^{10}\;M_\odot$) have compact nuclear star
clusters instead of BHs (e.g., Carollo et al. 1998; Matthews et al.
1999; B\"oker et al. 2002; 2004). Massive galaxies can have dense
nuclear clusters too (Seth et al. 2008). These clusters have about the
same correlations with bulge mass and velocity dispersion as the BHs
(Wehner \& Harris 2006; Rossa et al. 2006; Graham \& Driver 2007;
Ferrarese et al. 2006; C\^ot\'e et al. 2006; Li et al. 2007a), but
somehow the dense gas, whether in disk clumps or in the nuclei
themselves, made stars instead of BHs, or they made stars that could
not coalesce into BHs. In our model, this difference is mostly the
result of a difference in disk-clump density, with high density clumps
more likely to form IMBHs inside their dense cores and low density
clumps forming only dense stellar clusters. Dense clusters should
migrate to the galaxy centers in the same way as BHs if the clusters do
not evaporate first. This is consistent with the observation that
nuclear clusters typically contain a range of ages and are on average
younger than their disks (Rossa et al. 2006; C\^ot\'e et al. 2006). We
suggest that galaxies with the highest disk gas fractions and the
highest disk turbulent speeds make the densest and most massive disk
clumps, and that these are the formation sites for IMBHs. Less extreme
disks, or later stages of the same disks, make only dense star clusters
in their disk clumps.

The observed correlation between galaxy mass and the presence or lack
of nuclear BHs then follows from the correlation between galaxy mass
and density: higher mass galaxies are generally denser and should form
denser clumps and IMBHs in those clumps, while lower mass galaxies have
lower density disks and should form lower-density clumps that do not
make IMBHs. Both types of clumps will be relatively massive compared to
their disks (because both types of galaxies presumably have high ratios
of turbulent to rotational speeds at early times) and therefore both
types of clumps will interact and migrate to the center as we simulate
here. However, only the dense clumps in massive galaxies will bring
IMBHs to the nucleus. It is also possible that nuclear BHs form earlier
in the life of a massive galaxy than nuclear star clusters form in the
life of a low-mass galaxies, because of the shorter dynamical time for
the massive galaxy. In that case, nuclear clusters could take much
longer to form than nuclear BHs, when measured in absolute time. Galaxy
interactions should also increase the turbulent speed and disk gas
density, promoting the formation of IMBHs. In that case, tidal torques
as well as clump interactions would bring clump stars and IMBHs into
the center.

Dense nuclear clusters would not be expected to form IMBHs in the same
way as dense young clusters if the nuclear clusters assembled from
in-spiraling evolved clusters, which have no massive stars anymore. Low
mass stars are small and have much smaller gravitational cross sections
than high mass stars, and mass segregation in a young cluster puts the
high mass stars close together where they can interact quickly.  Thus
it is reasonable that young clusters in dense disk clumps make IMBHs,
which spiral in to the center to make nuclear BHs, while neither young
clusters in low-density disks nor evolved cluster debris in galactic
nuclei can make BHs by the same mechanism, which is direct stellar
collisions.

Our model applies primarily to BH formation at high redshift, which
occurs in galaxies that should be gas-rich and highly turbulent -- the
two primary ingredients for the evolution found here. Our model also
forms BHs rapidly and in a gas-rich environment, which triggers an
intense starburst. These aspects of BH formation are in agreement with
observations (e.g., Alexander et al. 2005; Escala 2006; Haiman,
Jimenez, \& Bernardi 2007). Supermassive BHs in $z\sim6$ quasars (Fan
2006) could also be formed by disk clump coalescence, because it takes
only a few disk rotations to bring the massive clumps to the center.
Galaxy interactions would have made clump migration and BH formation
even faster. However, our process would not apply to a starless central
region, such as that found by Walter et al. (2004) in a $z=6.42$ QSO;
gas-related accretion during mergers would be preferred there (e.g., Li
et al. 2007b). Our mechanism also cannot form nuclear BHs without
bulges, as observed in some galaxies (e.g., Filippenko \& Ho 2003).
Such BHs might have formed in nuclear clusters by stellar coalescence.

Our model predicts IMBH activity in high-redshift disk clumps. Such
activity may include X-ray, jet, and radio emission if the
self-absorption is not too large. We also predict that lower density
and more quiescent disks should make smaller disk clumps and possibly
no IMBHs, which would be replaced by clusters with too little density
for stellar coalescence. Clumpy galaxies at $z>1$ outnumber starburst
spirals and ellipticals by a factor 2 in the UDF (Elmegreen et al.
2007a). Given the short timescale of the clumpy phase in our model
($\sim$1~Gyr), all present-day early-type disk galaxies would seem to
have gone through a clump-cluster or chain-galaxy phase, forming their
bulges this way or adding to a smaller bulge formed earlier by
primordial galaxy mergers. Black hole growth by gas accretion should
follow their formation by IMBH coalescence. AGN feedback effects might
be important during the gas-accretion stage.

\acknowledgments

Numerical simulations were carried out on the NEC-SX8R vector computer
at CEA/CCRT. D.M.E. thanks Vassar College for publication support.
Helpful comments by the referee are appreciated.

{}

\clearpage

\begin{table}
\centering
%\begin{center}
\caption{Black Hole Migration and Final Bulge Properties}
\begin{tabular}{ccccccccccccc}
\tableline\tableline
Run & $N_i$ & $N_{250}$ & $N_{500}$ & $N_{1000}$ & $f_{M,500}$ & $f_{M,1000}$ & $M_{Bulge}$          & $M_{BH}$            & B/T & BH/B & $\sigma_{bulge}$\\
    &       &           &           &            &             &              &$\times10^9\;M_\odot$&$\times10^6\;M_\odot$&    &$(\times10^{-4})$& km s$^{-1}$\\
 \tableline
 0 & 6 & 3 & 4 & 5  & 0.69 & 0.93 &21    &11.3 & 0.30 & 5.4 &121\\
 1 & 7 & 3 & 7 & 7  & 1.00 & 1.00 &14.7  &9.7  & 0.21 & 6.6 &117\\
 2 & 7 & 2 & 5 & 6  & 0.78 & 0.92 &13.3  &6.4  & 0.19 & 4.8 &108\\
 3 & 7 & 3 & 6 & 7  & 0.86 & 1.00 &23.1  &16.8 & 0.33 & 7.3 &146\\
 4 & 6 & 2 & 3 & 5  & 0.61 & 0.83 &9.8   &3.2  & 0.14 & 3.3 &87\\
 5 & 6 & 2 & 5 & 5  & 0.91 & 0.91 &22.4  &19.2 & 0.32 & 8.6 &184\\
 6 & 5 & 3 & 5 & 5  & 1.00 & 1.00 &25.2  &22.2 & 0.36 & 8.8 &178\\
 7 & 8 & 2 & 5 & 6  & 0.86 & 0.91 &8.4   &3.8  & 0.12 & 4.5 &91 \medskip\\
0N & 8 & 5 & 8 & 8  & 1.00  & 1.00 &19.6  &16.5 & 0.28 & 8.4 &163\\
1N & 6 & 3 & 6 & 6  & 1.00 & 1.00 &12.6  &7.8  & 0.18 & 6.2 &118\\
2N & 7 & 3 & 6 & 6  & 0.87 & 0.87 &14.7  &8.3  & 0.21 & 5.7 &122\medskip\\
0B & 6 & 2 & 4 & 5  & 0.74 & 0.88 &24.6 &18.7 & 0.32 & 7.6 &145\\
1B & 7 & 4 & 6 & 7  & 0.91 & 1.00 &20.8 &19.8 & 0.27 & 9.5 &125\\
2B & 6 & 3 & 5 & 6  & 0.84 & 1.00 &19.3 &14.4 & 0.25 & 7.5 &109\medskip\\
\tableline
\end{tabular}
\tablecomments{\\
1. Run ID. Runs with N start with a NFW halo profile, runs with B start with an initial bulge.\\
2. Total number of black holes formed.\\
3. Number of black holes within 250 pc of the center at the end of the run (1Gyr).\\
4. Number of black holes within 500 pc at the end.\\
5. Number of black holes within 1 kpc at the end.\\
6. Fractional mass of all black holes that have reached the central 500 pc.\\
7. Same for the central 1000 pc.\\
8. Bulge mass (measured as in BEE07)\\
9. Final black hole mass within 500 pc\\
10. Final bulge-to-total mass ratio, not including dark matter in the total.\\
10. Final nuclear BH to bulge mass ratio, assuming that the IMBHs that
reach 500 pc eventually merge. \\
11. Bulge central line-of-sight velocity dispersion averaged from 50
projections uniformly distributed over the sine of the inclination
angle.}
\end{table}

\clearpage
%fig1
\begin{figure}
\centering
\includegraphics[width=6in]{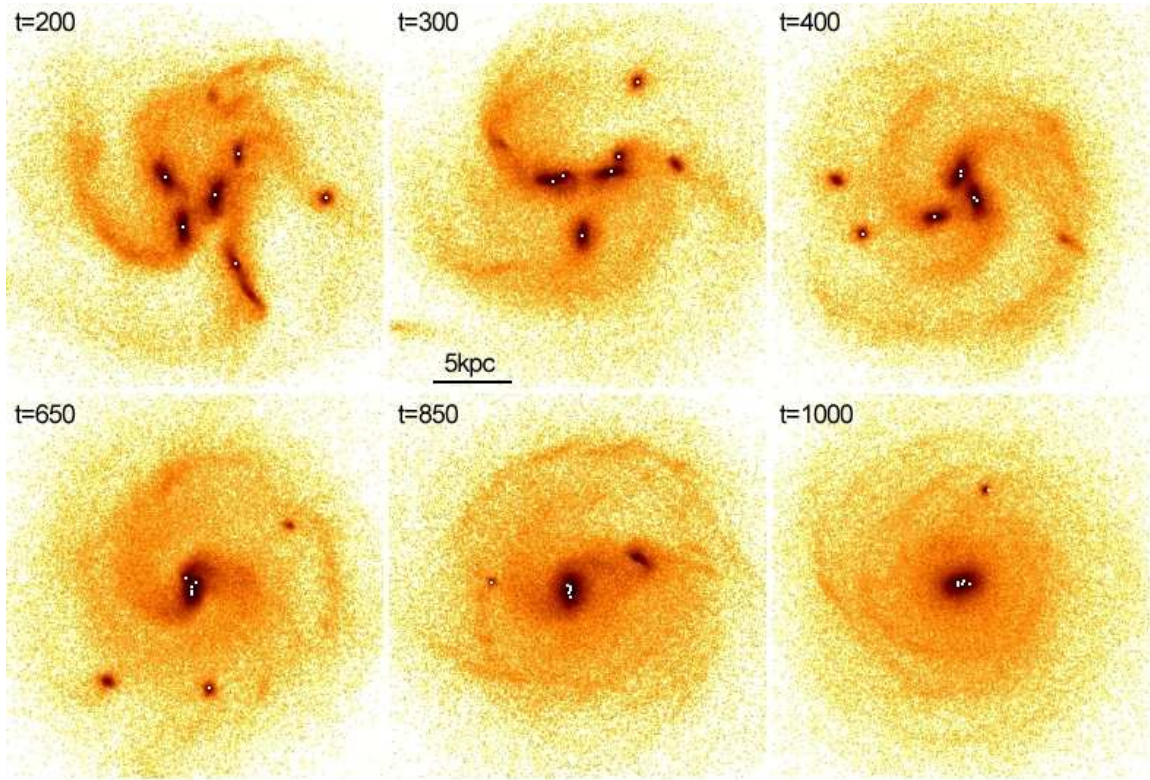}
\caption{Face-on snapshots of the disk mass density (gas and stars) for
run~0. White dots correspond to the positions of BH particles added at
the density peak of each clump when the mass fraction in the clumps
reaches its maximum. Time is in Myr. The BHs reach the center of the
galaxy as a result of clump interactions. }\label{fig:run0}\end{figure}

%fig2
\begin{figure}
\centering
\includegraphics[width=6in]{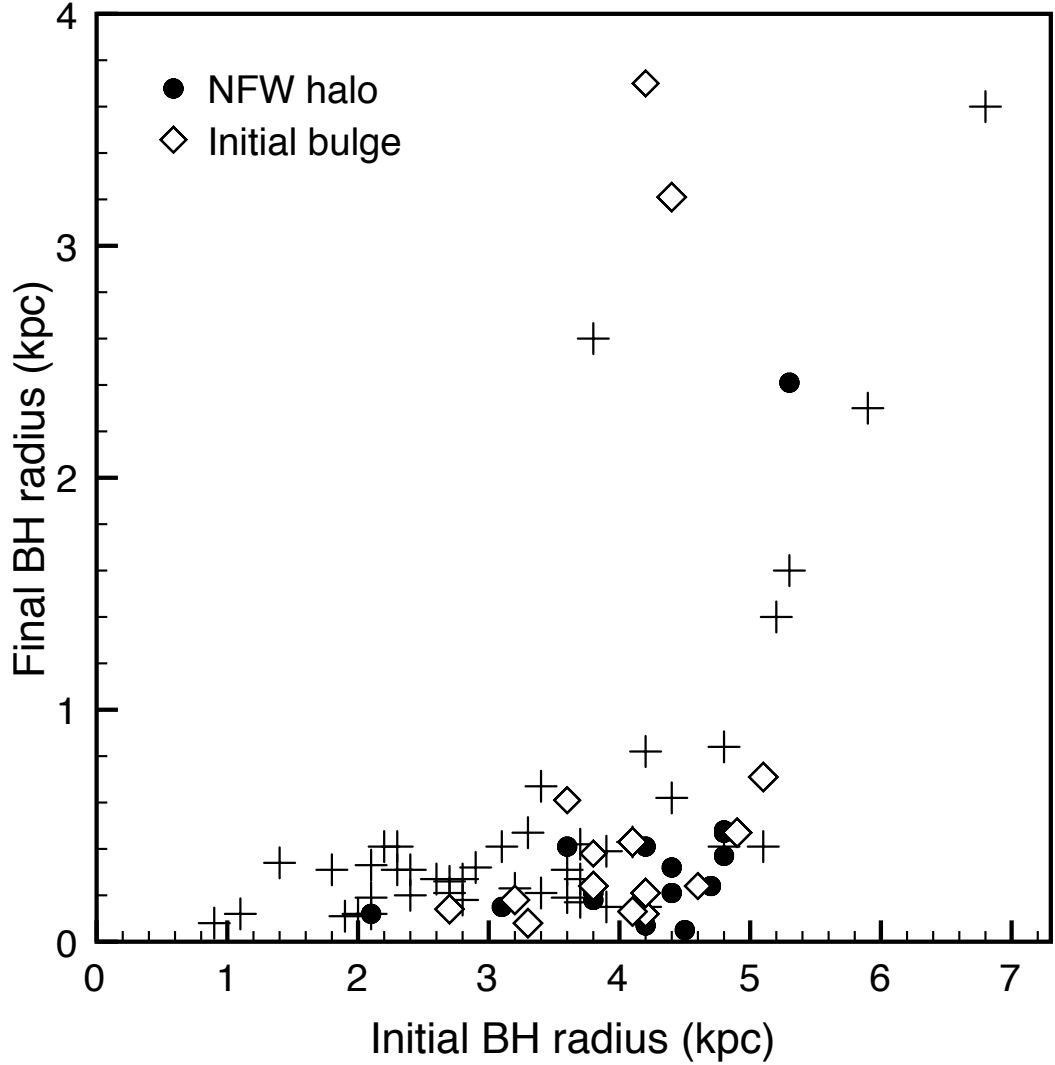}
\caption{Initial and final radius for each black hole particle for the
whole model sample. The initial radius is when the BH particle is
created, the final radius is after 1~Gyr. Plus signs correspond to
runs~0 to 7, circles to runs 0N, 1N and 2N (NFW halo) and diamonds to
runs 0B, 1B and 2B (models with an intial bulge). BHs that begin
further out in the disk are less likely to reach the
center.}\label{fig:BHrad}\end{figure}

%fig3
\begin{figure}
\centering
\includegraphics[width=6in]{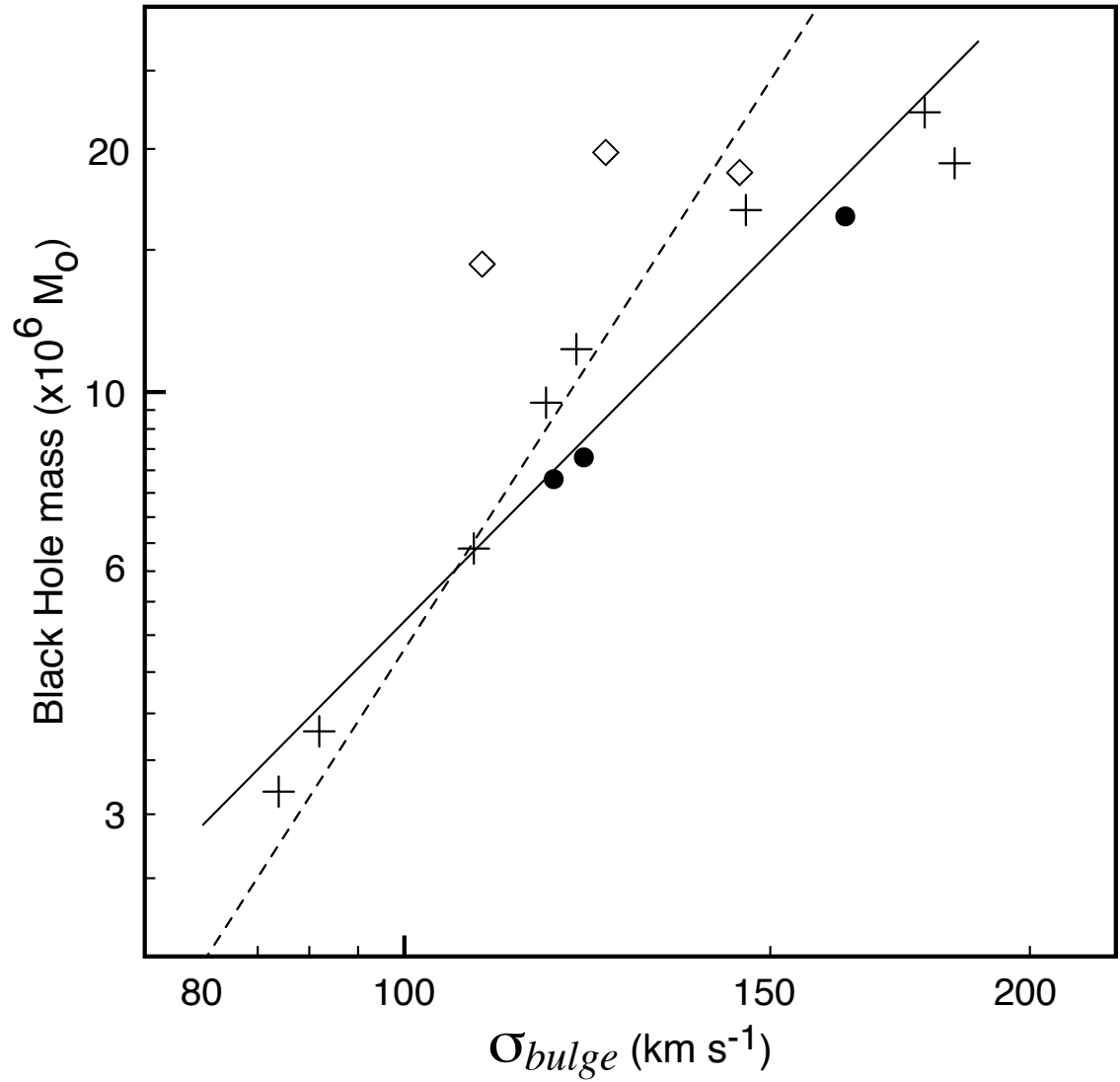}
\caption{The central BH mass (IMBH mass migrated within 500 pc) is
plotted versus the line-of-sight velocity dispersion of the bulge,
which is an average over 50 projections uniformly distributed over the
sine of the inclination angle. Each point is a different run, with the
same symbols as Figure~\ref{fig:BHrad}. The dashed line has a slope of
4, similar to the observations, and the solid line has a slope of 2 for
comparison. }\label{fig:disp}\end{figure}

\end{document}